\documentclass[12pt]{article}
\usepackage{epsfig,subfigure}
\usepackage{amssymb,amsmath,cite}

\begin{document}

\begin{titlepage}
\begin{flushright}
UCB-PTH-05/18\\
LBNL-57913\\
\end{flushright}

\vspace{15pt}

\begin{center}

{\huge\bf {Discretizing Gravity in Warped Spacetime}}

\vspace{15pt}

{Lisa Randall$^a$\footnote{randall@physics.harvard.edu},
Matthew D. Schwartz$^b$\footnote{mdschwartz@lbl.gov},
and Shiyamala Thambyahpillai$^a$\footnote{thamby@physics.harvard.edu}\\
}
\vspace{7pt}
{\small

${}^a$
{\it{Department of Physics, Harvard University \\
Cambridge, MA 02138\\}
}
\vspace{4pt}
${}^b$ 
{\it{Department of Physics, University of California, Berkeley,\\
      and\\
      Theoretical Physics Group, Lawrence Berkeley National Laboratory,\\
      Berkeley, CA 94720, USA\\}
}
}
\end{center}

\begin{abstract}
We investigate the discretized version of the compact Randall-Sundrum
model. By studying the mass eigenstates of the lattice theory, we
demonstrate that for warped space, unlike for flat space,
the strong coupling scale does not depend on the IR scale and 
lattice size. However, strong coupling does prevent us from
taking the continuum limit of the lattice theory.
Nonetheless, the lattice theory works in the manifestly
holographic regime and successfully reproduces the most
significant features of the warped theory. It is even in some
respects better than the KK theory, which must be carefully
regulated to obtain the correct physical results.
Because it is easier to
construct lattice theories than to find exact solutions to GR, we
expect lattice gravity to be a useful tool for exploring field
theory in curved space.
\end{abstract}

\end{titlepage}

\section{Introduction}
Warped geometries, such as the background used in Randall-Sundrum
model~\cite{rs1,rs2}, have provided many insights into general relativity and
holography. However, it is very difficult to find exact solutions
to Einstein's equations, so only a handful of warped geometries
are known. It would be useful to have a tool for constructing
theories  that reproduce the features of the warped geometries
without having to find and stabilize an appropriate gravitational
source. Discrete gravitational dimensions could be such a tool~\cite{orig,d1,d2,d3}. If
we work in the effective field theory framework with a cutoff,
 we may be able  
to learn a lot about general realtivity without needing exact solutions, as long
as the discrete theory can be trusted. Of course, this will
not tell us about the energy-momentum tensor to generate such a background,
but we can use the discrete model as a tool for investigating stability
of the system, the existence of ghosts, and the strong coupling scale,
for example. In this paper, we study the
the discretization of a single extra dimension in which we impose
 the exponential warp factor of the Randall-Sundrum model.
Although previous work discussed discrete nongravitational extra
dimensions for 
RS~\cite{rsw, Abe:2002rj,falk,Katz:2004qa,Bhattacharya:2005xa,Andreev:2004sy}, 
so far gravity has not
been included in the discretized model. We will find that many of
the features of the continuum gravitational theory are correctly
reproduced, and that some of the problems with flat space lattice
gravity are ameliorated.

The discretized model has a number of sites, each of which has a
separate four-dimensional  metric, and, in the minimal case, has
only nearest neighbor links. In the case of flat space, it has
been shown that the there is a limit to how small the lattice
spacing can be that depends on the overall size of the lattice.
This implies that there is no good continuum limit and it is
impossible to reproduce 5D gravity all the way up to the five
dimensional Planck scale. The absence of the continuum limit in
the flat case could be connected to the absence of a simple local
holographic description. It seems reasonable that a
four-dimensional discretized theory should exist for the warped
case, which does have a dual lower-dimensional description.

The flat space strong coupling problems result from two properties
of the  Kaluza-Klein spectrum: there is a very light mode and that
mode couples equally over the entire theory space. Both of these
properties are the result of the discretized kinetic term, which
introduces large mixing between the additional scalars that were
originally associated with a single site. The mixing leads to a
light highly delocalized mode, whose strong interactions
invalidate a low-energy description at a strong coupling scale
that lies well below the top of the KK tower. This is a less
severe problem in the warped geometry, because the strong
warp-factor dependence in the kinetic term keeps  the modes
localized on only a few sites. The absence of strong mixing means
that the localized modes are not lighter than you would naively
expect.

However, the warp factor only protects against mixing of modes
that are localized on sites further apart than the AdS curvature
scale. When the theory is discretized on smaller scales, the
theory resembles flat space. Within the AdS curvature scale, modes
do mix and get delocalized. So although it is possible to
reproduce many of the features of the continuum theory with the
discretized version, it is still not possible to achieve the true
continuum limit. Nevertheless, we find that the dangerous UV/IR
problem is absent: the lattice is self-consistent for any number
of sites.

\section{Set-up}
We will be concerned with the Poincare patch of $\mathrm{AdS}_5$, which is
described by the metric
\begin{equation}
  d s^2 = e^{- 2 k y} g_{\mu \nu} ( x, y ) d x^{\mu} d x^{\nu} + d y^2 .
\end{equation}
As in RS, we impose orbifold boundary conditions at $y = 0$ and $y = \pi R$.
The 5D gravitational Lagrangian in this background is
\begin{equation}
  \mathcal{L =} M^3_5 \sqrt{g_5} ( R_5 - 12 k^2 )
\end{equation}
\begin{equation}
  = M^3_5 \sqrt{g}  \left[ e^{- 2 k y} R_4 - 12 k^2 + \partial_y ( e^{- 2 k y}
  g_{\mu \nu} ) ( g^{\mu \rho} g^{\nu \sigma} - g^{\mu \nu} g^{\rho \sigma} )
  \partial_y ( e^{- 2 k y} g_{\rho \sigma} )  \right] .
\end{equation}
We are interested in linear fluctuations so we expand $g_{\mu \nu} = \eta_{\mu
\nu} + h_{\mu \nu}$. Then, the terms quadratic in $h$, after some integration
by parts, are
\begin{equation}
  \mathcal{L} = \frac{1}{4} M^3_5 \left\{ e^{- 2 k y} \left[ h_{\mu \nu} \Box
  h_{\mu \nu} + 2 h_{\mu \nu, \alpha}^2 - 2 h_{\mu \nu, \mu} h_{, \mu} + h_{,
  \mu}^2 \right] + e^{- 4 k y} \left[ ( \partial_y h_{\mu \nu} )^2 - (
  \partial_y h )^2 \right] \right\}\label{contlag} .
\end{equation}
To discretize the fifth dimension, we choose points evenly spaced in the
$y$-coordinate:
\begin{equation}
  y_j = j a, \quad j = 0 \cdots N ,
\end{equation}
where $N$ is the number of lattice sites and $a = R/N$ is the lattice
spacing. This is not the only possibility, and discretizing in a different
coordinate may result in a drastically different lattice theory. However, $y$
is in some sense the natural choice, since it is a geodesic coordinate for the
Poincare patch (unlike, say, the conformal coordinate $z$). But more
importantly, we can justify using $y$ \textit{a posteriori} because this
discretization will lead to a lattice theory with some of the holographic
features of AdS that we expect to see. We replace derivatives in the $y$
direction by differences. For example
\begin{equation}
  e^{- 4 k y} ( \partial_y h_{\mu \nu} )^2 \rightarrow \frac{1}{a^2} e^{- 4 k
  j a} ( h_{\mu \nu}^{j + 1} - h_{\mu \nu}^j )^2 .
\end{equation}
This brings out additional ambiguities related to the evaluation of the
warping prefactor, and to boundary terms which appear if we integrate by parts
before discretizing. However, again, our choices will be justified \textit{a
posteriori}. In any case, if the lattice theory is to be trusted, we should
expect that these ambiguities are irrelevant in the continuum limit, which we
discuss below.

Thus we get
\begin{equation}
  \mathcal{L} = \frac{1}{4} M^2 \sum_j e^{- 2 k a j} h_{\mu \nu}^j \Box h_{\mu
  \nu}^j + e^{- 4 k a j} \left[ \frac{1}{a^2} ( h_{\mu \nu}^{j + 1} - h_{\mu
  \nu}^j )^2 - \frac{1}{a^2} ( h^{j + 1} - h^j )^2 \right] ,
\end{equation}
with $M=\sqrt{M_5^3 a}$ the effective 4D Planck scale on the sites.
Going to canonical normalization
\begin{equation}
  \widehat{h_j} = M e^{- k a j } h_j
\end{equation}
we get a standard lattice action
\begin{equation}
  \mathcal{L} = \frac{1}{4} \sum_j \hat{h}_{\mu \nu}^j \Box \hat{h}_{\mu
  \nu}^j + M_{i j} ( \hat{h}^i_{\mu \nu} \hat{h}^j_{\mu \nu} - \hat{h}^i
  \hat{h}^j )
\end{equation}
with mass matrix
\begin{equation}
  M_{i j} = \frac{1}{a^2} e^{- 2 k a ( j - 1 )} \left[ ( e^{2 k a} + e^{- 2
  k a} ) \delta_{i j} - e^{  k a} \delta_{i + 1, j} - e^{- k a} \delta_{i -
  1, j} \right] .
\end{equation}
Note that this is essentially the same mass matrix as for a gauge 
boson~\cite{rsw,falk,Katz:2004qa,Bhattacharya:2005xa}.

At this point it is helpful to think about the values for $a,
k,$and $M$ that we would like to study, which are determined by
the continuum limit in which we are interested. As in flat space,
we would like to know whether we can have $N \rightarrow \infty$
in such a way that the lattice matches the continuum between any
two scales $\Lambda_{\mathrm{UV}}$ and $\Lambda_{\mathrm{IR}}$. In
the case of the compact Randall-Sundrum model relevant for the
standard model, which has $k \sim M$ and $R \sim 30 M^{-1}$, the
limitation $a > M^{- 1}$ implies that $N = R / a < 30$ in this
particular case. This is simply because in RS1 the fifth dimension
is only about 30 Planck units wide. Nonetheless, discretization on
the scale of $M^{-1}$ would be sufficient in principle to achieve
a theory that is valid up to the scale $M$, the same scale as the
continuum limit.

 In flat space, large $N$ allows you to
make a theory that is valid deeper in the infrared. This is also
true for  the RS model, but only by taking a larger ``volume" (that
is  \emph{R}). In this case,  $\Lambda_{\mathrm{IR}}$ would be
smaller. We will allow $k$, $N$, and $a$ to be free parameters,
with the understanding that $a$ will always be greater than
$M^{-1}$ and $N$ will be taken as large as necessary to achieve
any desired IR cutoff.

\section{Mass eigenstates}
In this section, we will develop insight into the mass eigenstates of
$M_{i j}$. We are working in the regime with $a \approx M^{- 1}$ and $k
\lesssim M$ so that the space is highly curved.
First, we make some quick observations about the mass matrix that will
uncover the important qualitative features of the warped lattice. We will afterwards
perform a more careful analysis.

\subsection{Rough, local flat space approximation \label{secrough}}
\begin{figure}[htbp]
\begin{center}
\epsfig{file=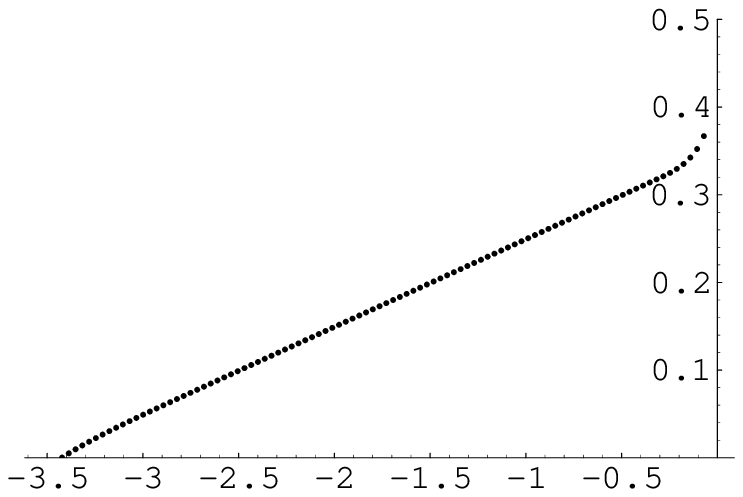,height=1.6in}
\epsfig{file=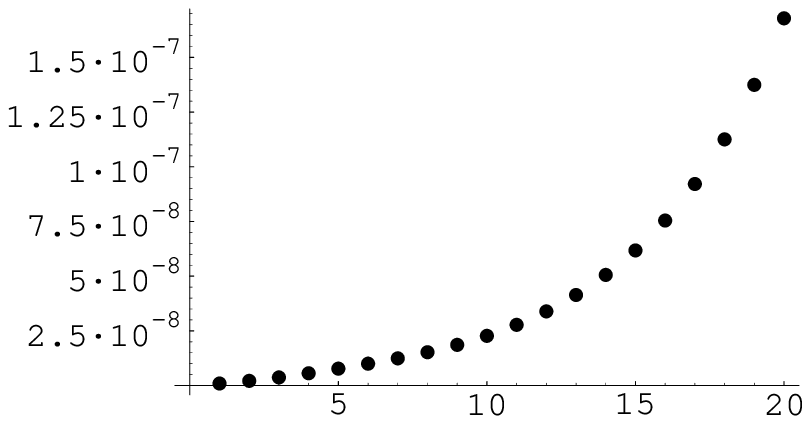,height=1.6in}
\caption{
The exact eigenvalues for $N=100$ and $ka = 0.1$.
On the left is a log plot, showing the geometric spacing.
On the right is a close up of the lowest 20 masses, in a non-log plot.
Note the linear spacing for the first
$1/(ka)=10$ modes.}
\label{figevals}
\end{center}
\end{figure}

Since $k a < 1$, the mass matrix is crudely approximated by
\begin{equation}
  M_{i j} \approx N_{i j} \approx \frac{1}{a^2} e^{- 2 k a j} \left[ 2
  \delta_{i j} - \delta_{i-1, j } - \delta_{i + 1, j} \right] \label{crude} .
\end{equation}
We have set $e^{k a}=1$ but not $e^{kaj}$, since $j$ may be large.
This mass
matrix scales with position and is almost diagonal. The only
non-diagonal entries are next to the diagonal, because of the
nearest-neighbor lattice structure we have assumed. But since the
position dependence of the warp factor is exponential, the mass
eigenvalues will be geometrically spaced:
\begin{equation}
  m_n \approx \frac{1}{a} e^{- k a n} .
\end{equation}
We confirm this geometric spacing numerically in Figure~\ref{figevals}.

This geometric spacing of the masses is very different from the linear
spacing of eigenvalues for the KK modes in RS. However, the linear
RS spacing is deceptive, since there are only of order $M/k$ relevant
at any site. The remaining modes are above the local cutoff, and
are in fact localized further in the UV.  That is, even though the classical
KK tower of RS has an infinite number of linearly spaced modes,
no calculation should ever involve
more than $M/k$ of them.
In fact, the linearized RS regime does appear, and the geometric spacing is actually
quite natural, reproducing the modes localized throughout the bulk with masses
determined by the local (warped) AdS scale.

To see this, consider the elements of this matrix
around $M_{j_n j_n}$ for some fixed site $j_n$. From Eq.~(\ref{crude}),
\begin{equation}
  M_{j_n - i, j_n - j} \approx \frac{1}{a^2} e^{- 2 k a j_n } e^{2 k a j}
  \left[ 2 \delta_{i j} - \delta_{i-1, j } - \delta_{i + 1, j} \right] .
\label{flatapp}
\end{equation}
For $j < ( k a )^{- 1}$ the $e^{2 k aj}$ term is approximately $1$. 
Then in this region the mass matrix
looks like flat space. So the eigenstates $H^{( j )}$ (for this
$j_n$) are roughly
\begin{equation}
  H^{( j )}_i \sim \sin ( \pi k a i j ), \quad j_n - \frac{1}{k a} < i < j_n +
  \frac{1}{k a} 
\end{equation}
with eigenvalues
\begin{equation}
  m_n \approx j ( k e^{- k a j_n } ) .
\end{equation}
For example, if we look at the lightest modes, with $j_n = N$ they
are linearly spaced just as in RS. This linear spacing breaks down
around $j\sim 1/(k a) \sim M/k$. This is exactly what we expect
because the $M / k$ mode has a mass of the local Planck scale. The
linear spacing for the low modes is evident in Figure~(\ref{figevals}).

In the approximation (\ref{flatapp}), the prefactor $e^{- 2 k aj_n}$ 
implies that there will be little mixing between modes from
an expansion around different sites $j_1$ and $j_2$ if $j_1 - j_2
> 1 / ( k a )$. The linear spacing is reconciled with the
geometric spacing which is transparent in the exponential
prefactor, because the linear spacing appears only when the
exponential is roughly constant. We will see that 
the lack of mixing between modes
localized more than of order $k^{-1}$ apart is the reason for the
larger regime of validity of the warped discretization, and
allows $N$ and $R$ to be taken as large as desired.

The interesting thing about this way of approximating the modes is
that it applies for modes centered around \textit{any} site $j_n$.
Of course, this is not surprising because of the conformal scaling
of RS as we move from the IR to the UV. But it is remarkable that
we can get information about a dynamical feature of RS, that the
relevant modes change with energy, from a fixed lattice.
Indeed, the modes near a site $j_1$ and those near a
site $j_2$ decouple, if $j_1 - j_2 > 1 / (ka)$.  This is
very different from what happens to the KK modes
 $\chi_n(y)$ of the continuum RS.
If we take two KK modes with masses matching the lattice modes, we
find that they do have significant overlap in even if their masses are very
different. Of course, the KK picture is not valid at all energies, so we should
never be considering two widely spaced modes at the same time. On the lattice
we can just use all the modes, and the widely spaced ones naturally decouple from
each other.
Although to use this rough flat-space approximation, 
we have to concentrate on a single site or a single mode, we will see that
with a more careful analysis of the lattice eigenstates, the same qualitative
features still hold.

\subsection{Improved solutions}
Even though this flat space approximation tells us most of what we
want to know about the lattice theory, it is instructive to have a
representation of the mass eigenstates in the region away from
where they have most of their support. This both justifies our
approximation, and will be used in the strong coupling
calculations below.

We want to develop a better understanding of the mass eigenstates of the
lattice theory. We are looking for the eigenstates of
\begin{equation}
  M_{i j} = \frac{1}{a^2} e^{- 2 k a ( j - 1 )} \left[ ( e^{2 k a} + e^{- 2
  k a} ) \delta_{i j} - e^{  k a} \delta_{i + 1, j} - e^{- k a} \delta_{i -
  1, j} \right] ,
\end{equation}
for $k a \lesssim 1$.
Let us define a matrix $\eta$ which goes from position eigenstates
to mass eigenstates
\begin{equation}
h_j = \eta_j^n H_n  \label{htobigH} .
\end{equation}
We have already observed from the flat space approximation,
that $\eta^n_j \sim sin( n j)$ for $|j-j_n| < 1/(ka)$.
Now we will confirm this and also approximate the $\eta^n_j$ in
other regions.

One approach is to use the   bulk Kaluza-Klein modes to guide the
search for the $H_n$. The KK equation following from the
Lagrangian (\ref{contlag}),
with $h(x,y)=\sum H_n(x) \chi_n(y)$, is
\begin{equation}
  \chi'' ( y ) - 4 k \chi' ( y ) + m_n^2 e^{2 k y} \chi ( y )
 = 0 \label{contkkeq}.
\end{equation}
It is not hard to see that the equation (\ref{contkkeq}) maps
directly onto the eigenvalue equation for $M_{i j}$.
The small parameter $k a$ is
the step size for the dimensionless continuum variable 
$\hat{y} = k y$. Then, expanding the derivatives, (\ref{contkkeq}) becomes
\begin{equation}
  \frac{1}{( k a )^2} [ \chi ( \hat{y} + k a ) - 2 \chi ( \hat{y} ) + \chi
  ( \hat{y} - k a ) ] - \frac{4 k}{( k a )} \frac{1}{2} \left[ \chi ( \hat{y}
  + k a ) - \chi ( \hat{y} - k a ) \right] + m_n^2 e^{2  \hat{y}} \chi (
  \hat{y} ) = 0.
\label{kkexp}
\end{equation}
For the first derivative, we have taken two steps to symmetrize with respect
to $\pm$. 

On the other hand, we can invert (\ref{htobigH})
\begin{equation}
H_n = \eta_n^j e^{-2k a j} h_j \label{Htoh} .
\end{equation}
We have used the fact that $H_n$ and $\hat{h}_j = e^{-kaj} h_j$ are canonically
normalized, so $\eta_n^j e^{-kaj}$ is unitary.
Then, hitting $H_n$ with $M_{i j}$, and projecting out the $h_j$ component we find
\begin{equation}
  \frac{1}{a^2} e^{- 2 k a ( j - 1 )} \left[ ( e^{2 k a} + e^{- 2 k a} )
  \eta^n_j - e^{-2k a} \eta^n_{j + 1}- e^{2k a} \eta^n_{j - 1} \right]
   - m_n^2 \eta^n_j = 0 \label{eveq}.
\end{equation}
For small $k a$ this becomes
\begin{equation}
  \frac{1}{a^2} \left[ \eta_{j + 1}^n - 2 \eta_j^n + \eta_{j - 1}^n
- 2 k a ( \eta_{j+1}^n - \eta_{j-1}^n ) \right] + e^{2 k a ( j - 1 )}
  m_n^2 \eta_j^n = 0,
\label{smallka}
\end{equation}
which is the same as (\ref{kkexp}) for $\eta_j^n = \chi_n (k a j)$.

In the continuum, the solutions to (\ref{contkkeq}) are KK modes:
\begin{equation}
  \chi_n ( y ) = e^{2 k y} \mathcal{J}_2 ( \frac{m_n}{k} e^{k y} ) .
\end{equation}
The corresponding discrete modes are
\begin{equation}
  \eta_j^n = e^{2 k a j} \mathcal{J}_2 ( \frac{m_n}{k} e^{k a j} ) .
\end{equation}
We cannot really manipulate these discrete Bessel functions, but we do expect
them to satisfy approximately the same relations as the continuum Bessel
functions. For example, the relation
\begin{equation}
  \chi_n' ( y ) = e^{3 k a j} \frac{m_n}{k} \mathcal{J}_1 ( \frac{m_n}{k} e^{k
  a j} )
\end{equation}
implies
\begin{equation}
  \eta_{j + 1}^n - \eta_j^n \approx k a \frac{m_n}{k} e^{3 k a j}
  \mathcal{J}_1 ( \frac{m_n}{k} e^{k a j} ) .
\end{equation}
Then it is rather trivial to show the modes are eigenstates -- they must be
because the equations match up.

But note that matching the continuum solutions onto the lattice
works because we are using only the first term in the Taylor series,
in going from (\ref{contkkeq}) to (\ref{kkexp}).
This is, in fact, all we can do, because we
have only included nearest-neighbor links in the discretization.
And that means that the solution will break down when the first
derivative is no longer a good approximation, i.e. when
\begin{equation}
  1 \approx k a \mathcal{J'} ( \frac{m_n}{k} e^{k y} ) \approx m_n a e^{k a
  j} .
\end{equation}
So these solutions should be good until $m_n e^{k a j} \approx 1 / a$. In fact,
the exact eigenstates of this matrix really  look like the continuum
KK Bessel functions only in this limited regime. The lattice does not just take
a selection of the KK modes, but presents a fundamentally different structure.
However, this structure actually reproduces the correct physics, as the heavier
KK modes in the continuum theory would be  cutoff by the local strong coupling scale.

The Bessel solutions apply for small $j <j_n$.
That is, they approximate the light KK modes at positions corresponding
to energies above their mass.
The Bessel functions peak and are localized when
\begin{equation}
m_n e^{k a j_n} = k ,
\end{equation}
which defines $j_n$.
So, the Bessel approximation is good through the localization region.
We also see that the width of the peak is
determined by the width of the exponential, that is
$\Delta j \approx 1 / (ka)$, so the region where the Bessel function peaks is the same as the region
where the flat space approximation applies.

In the flat space region, we can take
\begin{equation}
\eta^n_j = e^{2 k a j} \sin(\frac{\pi}{3} j) \sim  e^{2 k a j_n} \sin(j) ,
\end{equation}
which satisfies the small $ka$ equation, following from (\ref{smallka}):
\begin{equation}
2 \eta^n_j- \eta^n_{j+1} -\eta^n_{j-1}= e^{2 k a j} (a m_n)^2 \eta^n_j ,
\end{equation}
with the eigenvalues $m_n \sim  (1/a)e^{- k a n}$. 

These are not
quite the same flat space solutions as the ones derived in the
previous subsection from looking at the modes around a particular site. 
The difference is that
in the rough approximation, the heavier modes have higher frequency,
because they are the excited states of a box around a particular site.
We have seen here that the actual eigenstates 
all have the same frequency because each
one is effectively the zero mode of a different box, centered around the
site where the mode is localized.
Nonetheless, the important point is that
modes of both approximations have the
same qualitative features: oscillating behavior with roughly
constant amplitude over around $1/(ka)$ sites.

Now we have solutions for small $j$, in the Bessel regime, and for $j \sim j_n$,
in the flat space regime. For large $j$, so that $m_n a e^{k a j} \gg 1$ the
solutions are
\begin{equation}
  \eta_j^n \approx ( - 1 )^j e^{- k a j^2} e^{- k a j} ( m_n a )^{- 2 j} .
\end{equation}
We can check
\begin{equation}
  \eta_{j + 1}^n = - e^{- 2 k a j} e^{- 2 k a} ( m_n a )^{- 2} \eta_j^n 
\end{equation}
\begin{equation}
  \eta_{j - 1}^n = - e^{2 k a j} ( m_n a )^2 \eta_j^n .
\end{equation}
So in this regime $\eta_{j - 1}^n \gg \eta_j^n \gg \eta_{j + 1}^n$ and thus
(\ref{eveq}) reduces to
\begin{equation}
  - \frac{1}{a^2} e^{- 2 k a j} \eta^n_{j - 1} - m_n^2 \eta^n_j = 0 ,
\end{equation}
which is satisfied by our ansatz.

In summary, the solutions for the eigenstates are (up to normalization)
\begin{equation}
  \eta^n_j \propto
\left\{ \begin{array}{ll}
    e^{2 k a j} \mathcal{J}_2 ( \frac{m_n}{k} e^{k a j} ),
\quad \quad \quad\quad \quad \quad &j < j_n\\
    \sin ( j ),
& j \sim j_n\\
    ( - 1 )^j e^{- k a j^2} e^{- k a j} ( m_n a )^{- 2 j},
& j > j_n
  \end{array} \right. . \label{bigapprox}
\end{equation}
To align and normalize
the solutions, note that the Bessel
solution is exponentially growing
\begin{equation}
  e^{2 k a j} \mathcal{J}_2 ( \frac{m_n}{k} e^{k a j} ) \approx (
  \frac{m_n}{k} )^2 e^{4 k a j} .
\end{equation}
And in the third, large $j$, regime, $\eta^n_j$ dies as $\exp ( - k a j^2 )$.
So in both of these regimes, the modes have most of their support towards the
middle. This means that the normalization is determined almost entirely from
the flat space approximation.

\begin{figure}[htbp]
\begin{center}
\epsfig{file=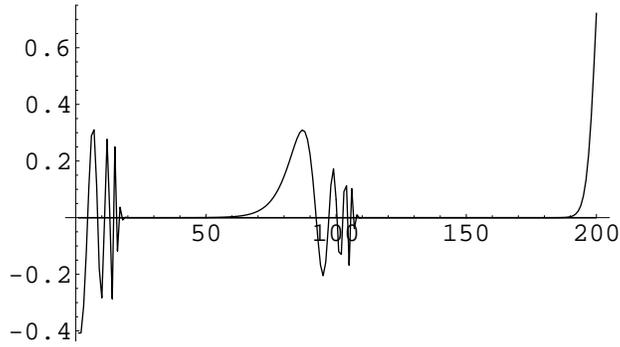,height=2in}
\caption{
Exact eigenvectors of the mass matrix as a function of site.
The numerical solutions for the 10th, 100th and $N$th eigenstates with $N=200$ 
and $1/(ka) = 10$ are shown.
Heavier modes are on the left. 
The $N$th mode is the lightest, and localized on the $N$th, IR, site.}
\label{figexact}
\end{center}
\end{figure}

\begin{figure}[htbp]
\begin{center}
\epsfig{file=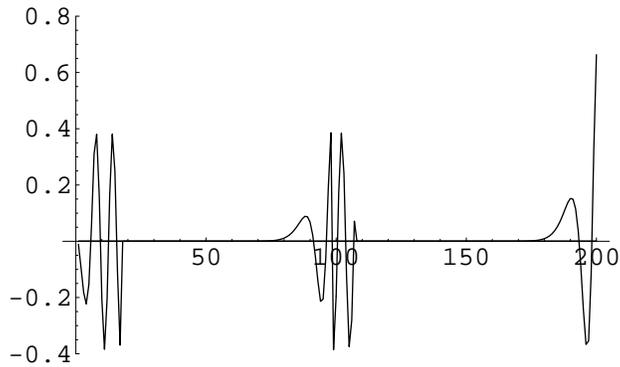,height=2in}
\caption{Approximate solutions, 
with the same parameters as in Figure \ref{figexact}}
\label{figapprox}
\end{center}
\end{figure}

Thus we are led back to our original guess. We only need the flat space
approximation. If we are looking near a site $j_n$ associated with a mode $n$
then
\begin{equation}
  \eta^m_j \approx \left\{ \begin{array}{ll}
    \sqrt{k a} \sin ( j ), &|j - j_n| < 1/(ka)\\
    0, & |j - j_n| > 1/(ka)
  \end{array} \right. \label{simpsum} .
\end{equation}
Some exact eigenvectors are plotted in Figure \ref{figexact}, and 
the approximation~(\ref{bigapprox}) is
shown in Figure \ref{figapprox}.

\section{Strong coupling}
Now that we understand the eigenstates of the lattice theory,
we can ask whether there is any limit in which the lattice is a good
approximation to the continuum. First, let us recall some difficulties
that are encountered in flat space~\cite{d2,d3}.

\subsection{Flat space}
To trust the predictions of the lattice theory, it is essential to
know the scale at which it becomes nonperturbative and the
effective theory breaks down. In a gravitational system, as in a
gauge theory, the easiest way to find this scale is to introduce
Goldstone bosons restoring a non-linear symmetry, \textit{a la}
CCWZ~\cite{Coleman:1969sm}. This procedure is explained in
detail in~\cite{d1}, and we will only briefly review it here.
For a massless graviton, this means restoring general coordinate
invariance with vector ($A_{\alpha}$) and scalar ($\phi$)
Goldstones:
\begin{equation}
  h_{\mu \nu} = g_{\mu \nu} - \eta_{\mu \nu} \rightarrow g_{\mu \nu} -
  \partial_{\mu} Y^{\alpha} \partial_{\nu} Y^{\beta} \eta_{\alpha \beta}
\end{equation}
\begin{equation}
  Y_{\alpha} \equiv A_{\alpha} + \phi_{, \alpha} .
\end{equation}
Then the Fierz-Pauli mass term generates a kinetic term  for the
scalar
\begin{equation}
  h_{\mu \nu}^2 - h^2 \rightarrow h_{\mu \nu}^2 - h^2 + \phi_{, \mu, \nu}^2 -
  ( \Box \phi )^2 + 2 \phi_{, \mu, \nu} ( h_{\mu \nu} - \eta_{\mu \nu} h ) + (
  \Box \phi )^3 + \cdots .
\end{equation}
The two four-derivative terms for $\phi$ cancel after integration by parts,
leaving a proper two-derivative kinetic term for $\phi$ after the $h, \phi$
kinetic matrix is diagonalized. The $\phi$ self-interactions are the strongest
and indicate where the theory breaks down. The scales
are
\begin{equation}
  \mathcal{L} = M_P^2 h \Box h + M_P^2 m^2 h^2 + M_P^2 m^2 \phi \Box h + M_P^2
  m^2 ( \Box \phi )^3 .
\end{equation}
So the canonically normalized fields are $h^c = M_P h$ and $\phi^c
= M_P m^2 \phi$ leading to
\begin{equation}
  \mathcal{L} = h^c \Box h^c + \phi^c \Box h^c + ( M_P m^4 )^{- 1} ( \Box \phi
  )^3 ,
\end{equation}
from which we read that the cutoff is $\Lambda = \Lambda_5 \equiv ( M_P m^4
)^{1 / 5}$.

On the flat space lattice, there are many massive gravitons. So we
introduce many Goldstones via $\Delta h_j \rightarrow \Delta h_j +
\Box\phi_j + \cdots$. Then
\begin{equation}
  \mathcal{L} = \sum_j M^2 h_{\mu \nu}^j \Box h_{\mu \nu}^j + \frac{M^2}{a^2}
  \left[ ( h_{\mu \nu}^{j + 1} - h_{\mu \nu}^j )^2 + ( h_{\mu \nu}^{j + 1} -
  h_{\mu \nu}^j ) \Box \phi^j + ( \Box \phi^j )^3 \right] .
\end{equation}
Barring any unforeseen cancellations the strong coupling scale is
determined by the interactions of the lightest mode. On the
lattice, the effective higher-dimensional modes (the KK modes) get
contributions from all the site modes, so there is a corresponding
$1 / N$ suppression of the interactions. For example, with $h_j =
\exp ( 2 \pi i j n / N ) H_n$
\begin{equation}
  \sum_j M^2 h_{\mu \nu}^j \Box h_{\mu \nu}^j \rightarrow
\sum_n N  M^2 H^n_{\mu \nu} \Box H_{\mu \nu}^n ,
\end{equation}
so the Planck scale associated with the sites is lower by a factor
of $\sqrt{N}$ than the Planck scale of the modes. For the gravity
case, there is an additional $N$-dependence for the scalar
longitudinal modes because they get their kinetic term from mixing
with differences. Then,
\begin{equation}
  \sum_j \frac{M^2}{a^2} ( h_{\mu \nu}^{j + 1} - h_{\mu \nu}^j )
  \Box \phi^j \rightarrow \sum_n N \frac{M^2}{a^2} \frac{1}{N} H \Box \Phi .
\end{equation}
Thus with $H^c = \sqrt{N} M  H$ we get $\Phi^c =  M / (
\sqrt{N} a^2 ) \Phi$. This means that the strongest interaction, from the
lightest mode, is
\begin{equation}
  \sum_j \frac{M^2}{a^2} ( \Box \phi^j )^3 \rightarrow N \frac{N^{3 / 2}
  a^4}{M} ( \Box \Phi^c_1 )^3 = \frac{1}{N M_P m_1^4} ( \Box \Phi^c_1
  )^3 ,
\end{equation}
where  $m_1 = 1 / N a$ and $M_P = \sqrt{N} M$. Thus the strong
interaction for the flat space lattice is at a scale
\begin{equation}
\Lambda_{\mathrm{flat}}= (M N^{-5/2} a^{-4})^{1/5}= ( N M_P m_1^4 )^{1 / 5}.
\end{equation}
Since this $M_P$ is the low energy Planck scale,
and $m_1$ is the physical mass, we can see that this scale is higher by a
factor of $N^{1 / 5}$ than that of a single massive graviton.

The fact that the strong coupling scale is heavier than the mass
of the light mode tells us that this is a good effective theory
for the light mode. But for a real lattice description of flat
space, this must be a good effective theory all the way up to the
5D Planck scale, $M_5^3 = M_P^2 R^{- 1}$. Since $R = m_1^{- 1}$ we
can write the cutoff as $\Lambda_{\mathrm{flat}} = ( M_5^3 R^{- 5}
a^{- 2} )^{1 / 10}$. Then for $\Lambda_{\mathrm{flat}} > M_5$ we
would need $M_5 a < ( R M_5 )^{- 5 / 2}$. But since $R M_5 \gg 1$
this means that $a \ll M_5^{- 1}$ -- the lattice spacing has to be
much smaller than the Planck length. But this does not make sense.
For the lattice to cut off the divergences from gravity, we must
take $a> M_5^{-1}$.

Another way to understand the difficulty is to observe that with
the lattice spacing at its limit, $a=M_5^{-1}$, we get
$\Lambda_{\mathrm{flat}} =\sqrt{M_5/R} \ll M_5$. So there will be
new effects at distances much much larger than the Planck length.
These non-local effects are most apparent if we work directly in
the continuum limit
\begin{equation}
  a^{- 2} ( \Delta h_j + \Box \phi_j + ( \Box \phi_j )^2 )^2 \rightarrow (
  \partial_y h )^2 + a^{- 1} ( \partial_y \phi ) ( \Box h ) + a^{- 2} ( \Box
  \phi )^3 + \cdots .
\end{equation}
An integration by parts has been performed on the middle term.
This shows that $\psi = \partial_y \phi$ is the propagating field,
while $\phi$ without derivatives is producing the strong
interactions. In terms of $\psi$ the Lagrangian contains
\begin{equation}
  \mathcal{L} = M_5^3 R h\Box h + M_5^3 R h \Box \psi + M_5^3 R a \left(
  \frac{\Box}{\partial_y} \psi \right)^3 + \cdots .
\end{equation}
The long wavelength modes with $\partial_y \sim R^{- 1}$ interact
at the scale $\Lambda_{\mathrm{flat}} = ( M_5^3 R^{- 5} a^{- 2}
)^{1 / 10}$ we derived above. But we can also now see that the
strong interactions are really non-local in the extra dimension.
Formally, as $a \to 0$ (or $N \rightarrow \infty $), the strong
coupling problem disappears, as it must if this lattice is to
classically reproduce the continuum. However, $\Lambda$ does not
grow with $N$ fast enough to ensure that the effective theory is
consistent.

\subsection{Warped space}
Now let us see how these observations change in the warped background. We
introduce Goldstones into the warped space Lagrangian in the usual way
\begin{equation}
  \mathcal{L} = \sum_j M^2 e^{- 2 k a j} h_{\mu \nu}^j \Box h_{\mu \nu}^j +
  \frac{M^2}{a^2} e^{- 4 k a j} \left[ ( h_{\mu \nu}^{j + 1} - h_{\mu \nu}^j
  )^2 + ( h_{\mu \nu}^{j + 1} - h_{\mu \nu}^j ) \Box \phi^j + ( \Box \phi^j
  )^3 \right] .
\end{equation}
The warp factor in front of the kinetic term tells us that the Planck scale on
a site is
\begin{equation}
  M_j = M e^{- k a j} .
\end{equation}
This is the warping we expect from the continuum. For the KK
modes, we saw that to a good approximation $1 / (k a)$ modes have
support at each site so we expect this parameter to play the role
of $N$ in the previous section. So the effective Planck scale for
the modes will be
\begin{equation}
  M_n = \frac{1}{\sqrt{k a}} M_{j_n}
\end{equation}
in agreement with the observation that $1 / (k a)$ in the warped
case should play the role of $N$.

In fact, because we are approximating the KK modes for warped space by
mapping to a position-dependent set of flat space solutions, we can simply use the flat
space results if we just complete the map. Looking at the Lagrangian, in the approximation
of Section \ref{secrough}, we see
that the warped space Lagrangian for the $1/(ka)$ modes around mode $n$ is equivalent
to a flat space lattice with
\begin{equation}
  M_{\mathrm{flat}} \rightarrow M_{j_n} = M e^{- k a j_n} \quad
  \frac{1}{a_{\mathrm{flat}}} \rightarrow \frac{1}{a} e^{- k a j_n} \quad
  N_{\mathrm{flat}} \rightarrow \frac{1}{k a} .
\end{equation}
Thus the strong coupling scale becomes
\begin{equation}
  \Lambda_{\mathrm{flat}} =
\left(  \frac{M_{\mathrm{flat}}}{N^{5/2}_{\mathrm{flat}}
     a_{\mathrm{flat}}^4} \right)^{1 / 5} \rightarrow
  \Lambda_{\mathrm{warp}} = \left( ( k a )^{5 / 2} M_{j_n} \frac{1}{a^4} e^{- 4 ka j_n}
\right)^{1/5} = ( \frac{1}{k a} M_n m_n^4 )^{1/5} \label{lwarp} ,
\end{equation}
where $m_n = k e^{-k a j_n}$ is the mass of the lowest
mode in the expansion around site $j$.

This result is superficially similar to that of flat space. The
strong coupling scale is a factor of $N_{flat} \sim 1 / (ka)$
above the strong coupling scale for a single massive graviton. But
there is a huge difference -- the strong coupling scale does not
depend on the size of the space we are latticizing. There is no
dependence on the total number of lattice sites $N$, or
equivalently on the IR scale $R= a e^{-kaN}$. In flat space, the
strong coupling was determined by the lightest mode, but in warped
space, it is determined by the lightest mode with support on the
site $j$. For a $j$ close to the UV brane, all the modes which
live there are much heavier than the $1/R$ mode which would
dominate if the space were flat.

However this is not the whole story.
In warped space, there is not a single strong coupling scale; the strong coupling scale depends on
the observer. In fact, there is an important difference between the scale associated with a particular
mode and the scale that an observer on a particular site would see.
For a mode of mass $m_n$, the
strong coupling scale is the usual
\begin{equation}
\Lambda_{\mathrm{mode}}\sim (M_n m_n^4)^{1/5} .
\end{equation}
This scale is heavier than $m_n$, so the mode
is weakly coupled at energies near its mass.
However, an observer at site $j$ would see $1/(ka)$ modes. In particular,
the mode of mass $m_n$ would be relevant even at energies as high as $m_n/(k a)$. When looking at the strong coupling scale on a site we must
 use $\Lambda_{\mathrm{warp}}$ of Equation (\ref{lwarp}). The relevant
Planck scale for the observer is $M_{j_n} = \sqrt{k a} M_n$. For example, with a lattice spacing
$1/a \sim M$ we would find that 
\begin{equation}
\Lambda_{\mathrm{site}} = \Lambda_{\mathrm{warp}}
= M_{j_n} \sqrt{\frac{k}{M}} < M_{j_n} .
\end{equation}
So, if we only talk about modes, there is no strong-coupling
problem in warped space. But observers localized at some position
in the bulk (that is, on a particular lattice site) must encounter
strong coupling before the local Planck scale. In asking about the
lattice theory, it makes sense to consider the site basis, since
otherwise the cut-off is determined solely by the IR regime.

We can understand these results in continuum language as well. In
the warped case,
\begin{equation}
  \mathcal{L} = e^{- 4 k a j } \frac{M^3_5}{a^2} ( \Delta h_{\mu \nu}^j + \Box \phi_j )^2
\rightarrow M^3_5 e^{- 4 k y} ( \partial_y h )^2
+ \frac{M^3_5}{a} e^{- 4 k  y} ( \Box \phi ) ( \partial_y h ) .
\end{equation}
When we integrate by parts, the $\partial_y$ term hits the warp
factor, so there are two pieces
\begin{equation}
  \frac{1}{a} e^{- 4 k y} ( \partial_y \phi - 4 k \phi ) ( \Box h )
\label{k} .
\end{equation}
Initially, we might expect that for the long wavelength modes,
with $\partial_y \sim 1/R$, the $k\phi$ piece will dominate and
prevent nonlocal effects. However, this would work only if there
were modes spread out over the whole space. In warped space, there
are not really any $1/R$ modes. The wavelengths in the extra
dimension are in fact limited by $k <
\partial_y < 1/a$ --  at each site, there are only $1/(ka)$ modes.
So $\partial_y \ge k$ and we can basically ignore the $k\phi$ term
in (\ref{k}).

Using $\hat{h} =exp(-ky) h$, and adding the $a/(ka)=1/k$ volume factor, the Lagrangian becomes
\begin{equation}
\mathcal{L}\supset \frac{M^3_5}{k} \hat{h}\Box\hat{h}
+\frac{M^3_5}{k}\frac{1}{a}e^{-3ky}\hat{h}\Box (\partial_y\phi )
+\frac{M^3_5}{k}\frac{1}{a^2}e^{-4ky}(\Box\phi)^3 .
\end{equation}
The canonical propagating field is then
\begin{equation}
\psi = \partial_y \phi^c =\frac{M^{3/2}_5}{k^{1/2}a}e^{-3ky}\partial_y\phi ,
\end{equation}
and the cubic scalar coupling becomes
\begin{equation}
\frac{M^3_5}{k a^2}  e^{-4ky}(\Box\phi)^3 \rightarrow
\frac{a k^{1/2}}{M^{3/2}_5 \partial_y^3} e^{5ky}(\Box\phi^c )^3 .
\end{equation}
For a mode of frequency $\partial_y \sim \omega$ this gives a strong coupling scale of
\begin{equation}
\Lambda^5 = \frac{M^{3/2}_5 \omega^3 }{a k^{1/2}} e^{-5ky} .
\end{equation}
If we look at the individual modes, then $\omega \sim 1/a$ 
(cf Eq (\ref{simpsum})), and so the mode scale is
\begin{equation}
\Lambda^5_{\mathrm{mode}} = \frac{M^{3/2}_5}{\sqrt{k}} \frac{1}{a^4}e^{-5ky} ,
\label{scale1}
\end{equation}
which is the same as $M_n m_n^4$ that we derived above. An
observer at position $y$ is sensitive to wavelengths as high as
$\omega \sim k$, which gives
\begin{equation}
\Lambda_{\mathrm{site}} = (M^3_5 k^5 a^{-2})^{1/10} e^{-ky} .
\end{equation}
This is the same as the flat space interaction scale
$\Lambda_{\mathrm{flat}} = (M_5^3 R^{-5} a^{-2} )^{1/10}$ with $k=R^{-1}$ the
playing the role of the IR cutoff for the warped space.

\section{Conclusion}
We have shown that a straightforward discretization of the warped AdS geometry 
produces a low-energy theory
which is valid above the scale of local curvature at any site.
This is in fact sufficient
for investigating many physical features, such as the
renormalization group behavior of the theory as
in~\cite{Pomarol:1999ad,Randall:2001gc,Randall:2001gb,Goldberger:2002cz}.
However, the discretization does not allow us to reach the UV cutoff
of the continuum theory. In the energy regime between the warped
AdS scale and the warped higher-dimensional Planck scale, the
theory acts like flat space and the same strong coupling problems
as in flat space appear. This is not surprising; at energies above
the local curvature, the theory approaches flat space, and therefore
we do not expect manifestly holographic behavior.
But it is extremely interesting
that some of the holographic behavior of warped space is also
manifested on the lattice; for example, we have found that
the mass eigenvalues are globally geometrically spaced, 
with only $M/k$ modes localized near any particular site.

An important distinction from the flat space theory is that 
we can take the large volume limit because the UV and IR cutoffs
 are independent.
In flat space, the UV cutoff goes down as the size of the space, $R$,
is increased. In curved space, the UV cutoff only depends on the
curvature scale, $k$; the cutoff in a particular region is completely
ignorant of the total volume, or equivalently, the total number of sites.
This cutoff is still below the local Planck scale, so
even in the warped case, the
lattice cannot be used as a regulator -- divergences must still be
cutoff by hand, or new physics must enter at a scale below the lattice
spacing.

As in any non-renormalizable theory with a dimensionful scale, there is a natural
limit to the lattice spacing, and thus a natural limit to the number of sites
on the lattice. In warped space, this is a particularly strong bound.
For example, in RS1~\cite{rs1}, the dimensionful scale is $M_P$, but the
size of the space is set by $k e^{-kR} = TeV$ and so
$N  < 30$.
Nonetheless, on a
larger warped space, approaching RS2~\cite{rs2}, $1/N$ effects can be parametrically
ignored. Note that it is precisely because we can take the large volume limit
with fixed lattice spacing that large $N$ is interesting.
In flat space, because the UV cutoff decreases as the volume grows,
 in the large
volume limit the cutoff goes to zero. In warped space,
 it remains above the local
curvature scale.

Because the discretization of RS is sufficiently well behaved, we expect
a similar discrete theory may be a useful tool for studying other geometries
that exhibit holographic behavior.
For example,
the metric for de Sitter space and some black holes can be written
 in a warp-factor
notation~\cite{Randall:2002tg}. Thus their discretizations 
should be similar and
may help unravel their holographic features.
Another interesting example is local localization~\cite{Karch:2001cw},
in which the warp factor decreases to a minimum and then grows again.
We expect that sites in the turnaround regime will
have a low cutoff. 
However, modes localized on these sites will have very little
support in the region where four-dimensional gravity applies.
There is no reason to expect the massive
mode of locally localized gravity that plays the role of the massive
graviton to cause problems.
Presumably locally localized gravity is far more general. The discrete
version of theories with non-monotonic warp factor could be a useful
tool for studying different 
examples of locally localized gravity, even with more than one extra dimension.

\section*{Acknowledgements}
We would like to thank J. Gallicchio and I. Yavin for discussing their
results~\cite{gy}. We would also like to thank Y. Shadmi and R.~Mahbubani 
for discussions. After this work was published, we were informed that a
related discussion can be found in \cite{Cognola:2003xr}.
The work of LR was supported in part by NSF Award PHY-0201124.

\end{document}